# Hunting for Monolayer Boron Nitride: Optical and Raman Signatures

R. V. Gorbachev[1*], I. Riaz[1], R. R. Nair[1], R. Jalil[1], L. Britnell[1], B. D. Belle[1], E. W. Hill[1], K. S. Novoselov[1], K. Watanabe[2], T. Taniguchi[2], A. K. Geim[1], P. Blake[1*]

[1]Manchester Centre for Mesoscience & Nanotechnology, University of Manchester, Manchester M13 9PL, UK
[2]National Institute for Materials Science, 1-1 Namiki, Tsukuba, 305-0044 Japan

blizza@gmail.com; peter@graphene.org

*We describe the identification of single- and few- layer boron nitride. Its optical contrast is much smaller than that of graphene but even monolayers are discernable by optimizing viewing conditions. Raman spectroscopy can be used to confirm BN monolayers. They exhibit an upshift in the fundamental Raman mode by up to 4 $cm^{-1}$. The number of layers in thicker crystals can be counted by exploiting an integer-step increase in the Raman intensity and optical contrast.*

Properties of few-nanometer-thick BN sheets (often referred to as few-layer BN) have been attracting steady interest over the last several years[1]. Although individual atomic planes of BN were also isolated[2] and investigated by transmission electron microscopy (TEM)[3-5] and atomic force microscopy (AFM)[6], interest in BN monolayers has been rather limited, especially, if compared with the interest generated by its "sister" material, graphene[7]. This can be attributed to 1) the lack of hexagonal boron nitride (hBN) crystals suitable for the mechanical cleavage approach[7] and 2) difficulties in isolating and finding sufficiently large BN monolayers. The situation is now changing rapidly due to the availability of hBN single crystals, which allow the cleavage of relatively large (~100 μm) and relatively thin (several nm) BN samples with an atomically flat surface.[6,8,9] Such crystals have been used as a thin top dielectric to gate graphene[9] and as an inert substrate for graphene devices, which allowed a significant improvement of their electronic quality,[8] unlike the earlier attempts with highly-oriented pyrolytic boron nitride (HOPBN)[10]. Most recently, it has been demonstrated that BN films with 2 to 5 layer thickness can also be obtained by epitaxial growth on copper and subsequent transfer onto a chosen substrate.[11] Particularly motivating is the emerging possibility to use BN as an ultra-thin insulator separating graphene layers. The layers could then be isolated electrically but would remain coupled electronically via Coulomb interaction, similar to the case of narrow-spaced quantum well heterostructures.[12] Such atomically thin BN-graphene heterostructures may allow a variety of new interaction phenomena including, for example, exciton condensation[13].

In the case of graphene, its mono-, bi- and few- layers are often identified by their optical contrast[14] and Raman signatures[15]. Little is known about these characteristics for the case of BN and, in the previous AFM and TEM studies,[2,5,6] one had to rely on finding atomically thin BN regions either randomly or close to edges of thick BN flakes. In this Letter, we report optical and Raman properties of mono- and few-layer BN obtained by micromechanical cleavage of high-quality hBN. Because of its zero opacity (the band gap is larger than 5eV),[1] atomically-thin BN exhibits little optical contrast, even if the interference enhancement using oxidized Si wafers is employed.[14,16] For the standard oxide thickness of ~300 nm $SiO_2$,[6,7] BN monolayers show white-light contrast of <1.5%, which makes them undetectable by the human eye.[17] Moreover, the contrast changes from positive to negative between red and blue parts of the spectrum, respectively, and goes through zero in green where eye sensitivity is maximum. We show that the use of thinner $SiO_2$ (≈80±10 nm) offers optimum visualization conditions with contrast of ~2.5% per layer, similar to that for graphene on transparent substrates using the light transmission mode. Mono- and bi- layers can also be identified by Raman spectroscopy due to shifts in position of the characteristic BN peak that is centered at ≈1366 $cm^{-1}$ in hBN crystals.[1] Monolayers exhibit sample-dependent blue shifts by up to 4 $cm^{-1}$. This is explained by a hardening of the $E_{2g}$ phonon mode due a slightly shorter B-N bond expected in isolated monolayers,[18] with further red shifts due to random strain induced probably during the cleavage. This strain effect dominates in bilayer, causing red shifts of the Raman peak by typically 1 to 2 $cm^{-1}$.

Atomically thin BN crystals were prepared by the standard cleavage procedures[2] and using hBN single crystals grown as described in refs. 19,20. It is important to note that previously we used HOPBN (*Momentive Performance Materials*) but could only obtain strongly terraced crystallites and no monolayers.[10] BN monolayers mentioned in ref. 2 were extracted from a powder (*Sigma-Aldrich*) and did not exceed a couple of microns in size because of the small size of initial flakes. Using hBN, we can now prepare few-layer samples larger than 100 μm, that is, comparable in size to our single crystals. Figure 1 shows examples of single- and few- layer BN on top of an oxidized Si wafer. The AFM images in Fig. 1 are to illustrate our identification of regions with different thickness.

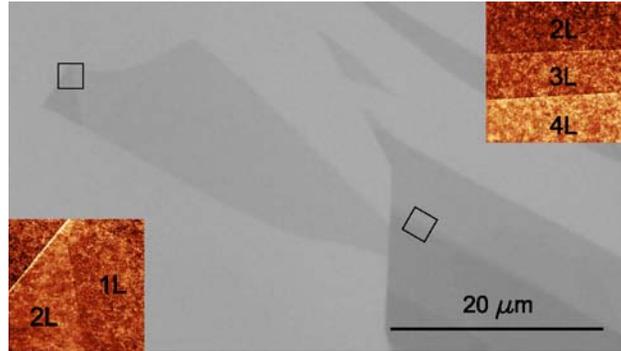

**FIG 1**. (Color online) Atomically thin BN on top of an oxidized Si wafer (290 nm of $SiO_2$) as seen in an optical microscope using a yellow filter (λ =590 nm). The central crystal is a monolayer. For legibility, the contrast is enhanced by a factor of 2. The insets show AFM images of the 3.5x3.5μm² regions indicated by the squares. The step height between the terraces in the images is ~4Å. BN crystals are usually lifted above the wafer by up to extra 10Å, which can be explained by the presence of a water or contamination layer.[2,6]

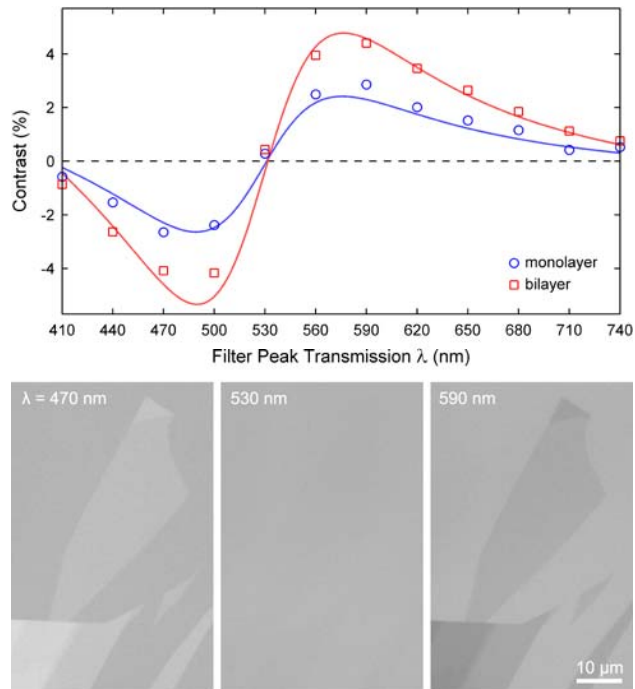

**FIG 2**. (Color online) Changes in the optical contrast with wavelength for mono- and bi- layer BN on top of a Si wafer (290 nm $SiO_2$). We used filters with a 10 nm bandwidth. The solid curves are the dependences expected for mono- and bi- layer BN. In the modeling, we have included the influence of a finite numerical aperture (NA).[21] For the used microscope objective (NA =0.8), we have integrated over angle assuming a Gaussian weight distribution of width $\theta_{NA}/3$ where $\theta_{NA}$ is the maximum acceptance angle of the objective lens.[22] The lower panels show examples of the BN visibility using different filters for the same sample as in Fig. 1. For legibility, the contrast in the images has been enhanced by a factor of 2.

Figure 2 shows variation of the contrast measured with respect to the bare wafer at different wavelengths λ. To this end, we have taken optical micrographs using illumination through narrow bandpass filters.[14] Representative images for 3 different λ are also presented in Fig. 2. One can see that the contrast is a nonmonotonic function of λ and changes its sign at ~530 nm (BN is darker than the substrate at long wavelengths and brighter at short ones). This is different from graphene, in which case the contrast is either positive or negligible.[14] With increasing the number of BN layers *N*, the contrast increases proportionally to *N*. To explain the measured λ dependence, we have used an analysis similar to that reported for graphene[14] and based on the matrix formalism of interference in thin film multilayers[16]. This requires the knowledge of the real n and imaginary k parts of the refractive index. We used spectroscopic ellipsometry for our hBN crystals and found $k \approx 0$ and $n \approx 2.2$ with a slight upshift for λ <500 nm. Assuming that optical properties of monolayers change little with respect to hBN, we obtain the dependences shown in Fig. 2. The theory accurately reproduces the observed contrast, including its reversal at 530 nm and the absolute value that is related to the extra interference path due to the presence of a transparent monolayer on top of $SiO_2$.

The developed theory allows us to predict at which $SiO_2$ thickness the optical contrast for BN monolayers would be maximal. Fig. 3 shows that this is expected for a thickness of 80±10 nm. In this case, the contrast remains relatively strong with the same sign over nearly the entire visible range. This prediction has been confirmed experimentally by imaging BN crystals on top of 90 nm $SiO_2$. We have found that the contrast reaches ~2.5% per layer already in white light (~3% with a green filter), and this is sufficient to hunt for and directly see BN monolayers in a microscope. Still note that it is much harder to find BN than graphene monolayers that give a contrast of ~10%[14].

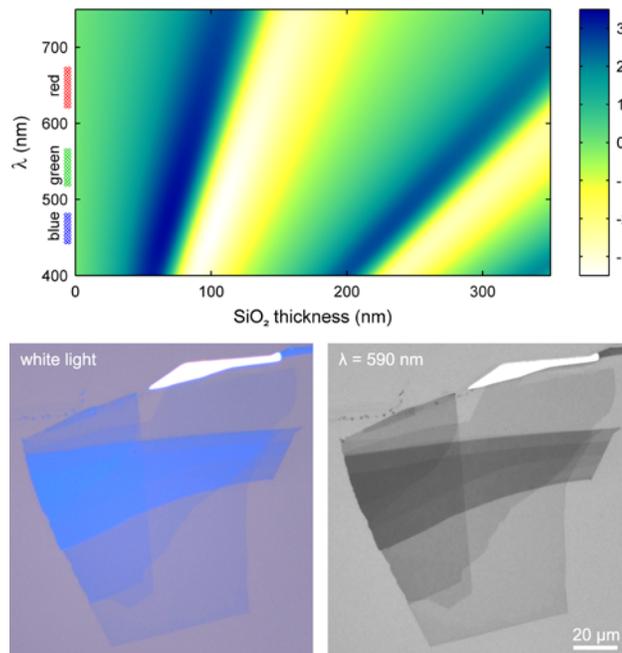

**FIG 3**. (Color online) Optical contrast due to monolayer BN for different λ and $SiO_2$ thicknesses (top). The plot is for the case of a typical high magnification objective (50X) with NA =0.8 but changes little for NA =0.7 or 0.9. The lower images show BN on top of a 90 nm $SiO_2$/Si wafer (the lower part is a monolayer). Similar to Figs. 1&2, the contrast is enhanced by a factor of 2.

The optical contrast increases in integer steps (that is, by a factor of *N* for *N*-layer BN) and this can be employed for search and identification of mono- and few- layers. However, let us warn that any contamination or a thin layer of water, which is believed to raise atomic crystals above Si wafers, can notably affect the measured contrast. This was previously observed for graphene[22] but the effect becomes much more important for BN because of its weaker contrast. In our experience, it is not unusual for monolayer BN to look like a bilayer. To avoid misidentification and obtain the correct contrast as reported above we annealed our samples at 150°C in vacuum. For this and other reasons, it is desirable to have another way of confirming BN thickness. Of course, AFM can be used to this end but it is a low throughput technique. For the case of graphene, Raman spectroscopy has proven to be an indispensible tool and, below, we show that it is also useful for identifying monolayer BN.

Figure 4a shows Raman spectra of mono-, bi- and tri- layer BN using a green laser with λ =514.5 nm. BN exhibits a characteristic peak that is due to the $E_{2g}$ phonon mode and analogous to the G peak in graphene.[1,18] In our hBN single crystals, the Raman peak occurs at ≈1366 cm$^{-1}$. One can see in Fig. 4a that the peak becomes progressively weaker as *N* decreases and, for monolayer BN, its intensity is ~50 times smaller than for graphene's G peak under the same measurement conditions. We have found that the integrated intensity $I_T$ for the BN peak is proportional to *N* with high accuracy for first several layers (inset in Fig. 4a). Accordingly, once a Raman spectrometer is calibrated for a given substrate, this can be exploited to distinguish between one, two and more BN layers.

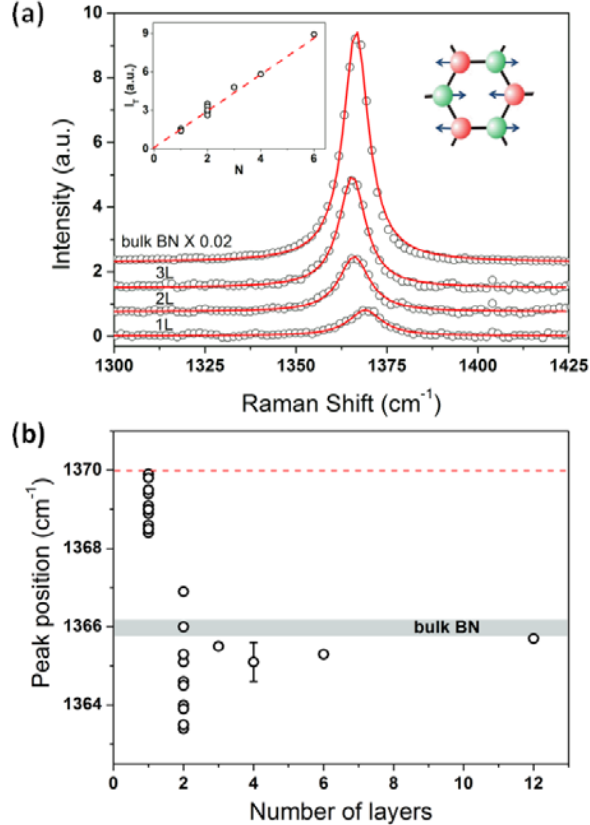

**FIG 4.** (Color online) (a) Raman spectra of atomically thin BN. The left inset show changes in integrated intensity $I_T$ with the number of layers *N*. The right picture illustrates the phonon mode responsible for the Raman peak. (b) Position of the Raman peak for different *N*. In mono- and bi-layer BN, the peak position is sample dependent and varies by as much as ± 2 cm$^{-1}$. The dashed line is the Raman shift predicted for monolayer BN[18]. The error bar indicates a typical accuracy of determining the peak position using our spectrometer.

In addition to its intensity proportional to *N*, we have found that the Raman peak is usually shifted upwards in monolayers and downwards in bilayers with respect to its position in bulk hBN (see Fig. 4b). Monolayers show relatively large shifts (typically, between 2 to 4 cm$^{-1}$), which vary from sample to sample. The maximum observed blue shift is in agreement with the theory expecting its value to be ≈4 cm$^{-1}$ for monolayers.[18] However, Fig. 4 also shows that mono- and bi- layers exhibit unexpectedly strong variations in the peak position whereas these are essentially absent for crystals thicker than 5 layers (not all data for thicker crystals are shown in Fig. 4). To find the origin of these changes, we used different laser powers and ruled out heating effects. We also measured the width of the Raman peaks. The HWHM varied between 10 and 12 cm$^{-1}$ for monolayers and was only marginally larger than the width in hBN (≈9 cm$^{-1}$). No apparent correlation between the width and peak position was found.

To explain the observed variations, we invoke strain that causes additional sample-dependent red shifts in the case of stretching. Indeed, for graphene, the analogous G peak is red-shifted by as much as ~20 cm$^{-1}$ per 1% of strain[23]. Strain-induced shifts in graphene deposited on a substrate are completely masked by doping effects[24] which often move the G peak by ~10 cm$^{-1}$. In the absence of such doping effects for insulating BN, strain is expected to become an important factor in determining the Raman peak position. The observed downshifts with respect to the intrinsic blue shift would then imply the stretching of BN monolayers by only a fraction of a percent, which is highly feasible. It seems pertinent to attribute the peak broadening to the same effect. Indeed, strain can also vary within the micron-sized laser spot as monolayers try following the substrate roughness[25]. This argument also applies for bilayers and can explain their random shifts and notably smaller broadening (HWHM of ~9 to 10 cm$^{-1}$). The maximum observed peak position for bilayers in Fig. 4b implies a small intrinsic blue shift of ~1 cm$^{-1}$. We are not aware of any theory for the intrinsic shift in BN bilayers.

In conclusion, BN mono- and bi-layers can be prepared and identified on top of an oxidized Si wafer using the same mechanical exfoliation technique as widely employed for the case of graphene. BN monolayers obtained from hBN crystals can be as big as samples of cleaved graphene and, therefore, should allow a variety of new experiments and proof-of-concept devices, beyond the previous studies by AFM and TEM. The search for atomically thin BN is more difficult than for graphene as the former does not absorb visible light and, therefore, gives rise only to the contrast due to changes in the optical path. Nevertheless, the use of thinner SiO$_2$ and/or narrow optical filters makes it possible to see even BN monolayers. To verify the number of layers, one can employ Raman spectroscopy. It allows the identification of monolayers by an upward shift in the Raman peak position. The shift depends on local strain and, therefore, is not as unambiguous as the Raman signatures for mono- and bi- layer graphene. The step-like increase in the Raman intensity can be used for further confirmation and for counting the number of layers. We believe that the provided analysis and the strategy for hunting for mono- and few-layer BN should facilitate further work on this interesting two-dimensional insulator.

*Acknowledgements* - The work was supported by EPSRC (UK). We thank Andrea Ferrari for many helpful discussions.